\documentclass[11pt]{article}
\usepackage{latexsym}
\usepackage{amsmath}
\usepackage{amsfonts}
\usepackage{dsfont}
\usepackage{verbatim}
\usepackage{hyperref}
\usepackage{subcaption}
\usepackage{amsmath,amssymb}
\usepackage{graphicx}
\usepackage{color}
\usepackage{enumerate}
\usepackage{booktabs}
\usepackage{multirow}
\usepackage{url}
\usepackage[shortlabels]{enumitem}
\usepackage{notoccite}
\usepackage[ruled,vlined]{algorithm2e}
\usepackage{xcolor,colortbl}
\usepackage{placeins}
\usepackage{authblk}
\usepackage{sidecap}

\title{Uncertainty-Error correlations in Evidential Deep Learning models for biomedical segmentation}
\author[1,2]{Hai Siong Tan\footnote{Corresponding Author. Email: haisiong.tan@gryphonai.com.sg}}
\author[3]{Kuancheng Wang}
\author[2]{Rafe Mcbeth}
\affil[1]{Gryphon Center for Artificial Intelligence and 
Theoretical Sciences, Singapore}
\affil[2]{University of Pennsylvania, Perelman School of Medicine, Department of Radiation Oncology, Philadelphia, USA}
\affil[3]{Georgia Institute of Technology, Atlanta, GA, USA}
\date{}                    
\def\be{\begin{equation}}
\def\ee{\end{equation}}
\def\bea{\begin{eqnarray}}
\def\eea{\end{eqnarray}}

\makeatletter
\renewcommand{\@algocf@capt@plain}{above}% formerly {bottom}
\makeatother

\begin{document}

\maketitle

\begin{center}
\textbf{Abstract}
\end{center}
In this work, we examine the 
effectiveness of an uncertainty quantification framework known as Evidential Deep Learning applied in the context of biomedical image segmentation.
This class of models involves assigning Dirichlet distributions as priors for segmentation labels, and enables a few distinct definitions of model uncertainties.
Using the cardiac and prostate MRI images available in the Medical Segmentation Decathlon for validation, we found that Evidential Deep Learning models with U-Net backbones generally yielded superior correlations between prediction errors and uncertainties relative to the conventional baseline equipped with Shannon entropy measure, Monte-Carlo Dropout and Deep Ensemble methods.  We also examined these models' effectiveness in active learning, finding that relative to the standard Shannon entropy-based sampling, they yielded higher point-biserial uncertainty-error correlations while attaining similar performances in Dice-S\o rensen coefficients. These superior features of EDL models render them well-suited for segmentation tasks that warrant a critical sensitivity in detecting large model errors. 
\\[15ex]
\textbf{Keywords}: uncertainty quantification, deep evidential learning, radiotherapy dose prediction 
\newpage
\tableofcontents

\section{Introduction}

This paper examines the 
effectiveness of an uncertainty quantification (UQ) framework known as Evidential Deep Learning (EDL) applied in the context of U-Net-based segmentation of biomedical images. In particular, we scrutinize the nature of uncertainty-error correlations inherited by this model, comparing them to two major categories of UQ frameworks in the current literature\footnote{A general survey can be found in \cite{abdar} whereas comprehensive recent reviews in the field of medical 
imaging can be found in \cite{lambert} and \cite{zouReview}. }:
(i) Monte-Carlo (MC) Dropout (ii) Deep Ensemble method.

The fundamental principle of EDL
lies in asserting a higher-order Bayesian prior over the probabilistic neural network output, embedding it within the loss function and subsequently generating uncertainty heatmaps upon completion of model training. There are two major variants of it proposed in the seminal works of Sensoy et al. in \cite{sensoy} and Amini et al. in \cite{amini}, with the former being more relevant for our work here. In EDL, these parameters are incarnated as part of the model output which are thus naturally obtained when training is completed. They are then used to furnish estimates
of model uncertainty. 
This is in contrast to MC Dropout and Deep Ensemble techniques. 
For MC Dropout \cite{gal}, one typically inserts stochastic dropout variables at various points of the neural networks, and after model is trained, a number of feedforward passes are performed to evaluate mean and variance of the output. The latter is then regarded as the model uncertainty. For Deep Ensemble method \cite{lak,Ganaie}, one typically sets up a list of models sharing the main network structure but differing in initial weight values and possibly model hyperparameters so that the final output is obtained as the average over this ensemble, and the uncertainty defined as the variance of the set of outputs.

In this paper, a primary focus lies in examining the degree to which uncertainty estimates are correlated with model prediction errors. 
If there exists a robust uncertainty-error correlation for the UQ model, it implies that one can leverage the uncertainty distribution
to discover potential regions of model errors, and thus enable efficient human expert intervention as part of a reliable A.I.-guided workflow. This feature can be crucial for certain real-life biomedical segmentation tasks that warrant a higher sensitivity in detecting model errors. For example, in the radiotherapy clinic, if the autosegmentation of a lesion target in CT images suffers from a serious error made by the model, it may lead to harmful radiation dose being deposited on organs-at-risk. An uncertainty-aware segmentation model yields uncertainty heatmaps that can warn the medical team against potential large model errors. This in turn relies on the premise that the model inherits a strong \emph{uncertainty-error correlation} -- which is the primary focus of our work here.

To our knowledge, applications of EDL in the context of biomedical segmentation have previously appeared in 
\cite{Jones,Hao,Shi,Zou}. However, none of these earlier works (and including others, e.g. \cite{Bao,stereo,molecules,Tong} devoted to applications outside the domain of segmentation) demonstrated any direct evidence that uncertainty estimates correlate well with the actual model errors characterizing predictions made on the testing dataset. It is thus unknown how different UQ models compare against one another in this aspect. Our work fills this gap. Using
two different MRI datasets drawn from the Medical Segmentation Decathlon \cite{Decathlon}, we furnished contextual definitive evidence that EDL models exhibit stronger uncertainty-error correlations for the segmentation task while attaining similar Dice coefficients compared to the more conventional MC Dropout and Deep Ensemble methods. We found that the uncertainty heatmaps generated by EDL models 
can act as useful visual aids towards identifying potential regions of model errors. Further, we found highly suggestive evidence of how EDL-based uncertainty sampling can furnish an excellent active learning method, attaining higher uncertainty-error correlations while converging to the same optimal Dice coefficient (associated with the full training set)
compared to Shannon entropy-based uncertainty sampling.

\section{Preliminaries}

Before elaborating on our model implementation, we would like to briefly review some fundamental aspects of the theoretical framework underlying
Evidential Deep Learning (EDL), in particular, with a focus on how it contains a few different notions of uncertainties; this has been missed by most of the works using EDL (e.g. \cite{Hao,Shi,Zou}).  

We can consider segmentation of an organ on some medical image as a set of Bernoulli trials defined separately for each pixel. For a binary class segmentation, let $p_2$ be the probability of the pixel belonging to the foreground (part of the organ) and $p_1 = 1- p_2$ that of it being part of the image background. A neural network adapted for this task will typically generate softmax final outputs corresponding to these probability parameters. 
A key feature of EDL is the notion of specifying a prior distribution for the Bernoulli or more generally the categorical distribution, i.e. that $p_1, p_2$ themselves are random variables following a `higher-order' distribution. In Bayesian language, the prior distribution is some 
assumed probability distribution before any observational evidence is taken into account.
A common prior or `evidential' distribution adopted for a general multi-class segmentation tasks in EDL is the Dirichlet distribution 
\be
\label{Diri}
D \left(\vec{p} | \vec{\alpha} \right)  =  \frac{1}{B(\vec{\alpha})} \prod_{i=1}^K p_i^{(\alpha_i - 1)}, \qquad
\sum_{i=1}^K p_i = 1, \qquad B(\vec{\alpha}) = 
\frac{\prod_{i=1}^K \Gamma (\alpha_i)}{\Gamma(\sum^K_{i=1} \alpha_i)},
\ee
where $K$ is the number of class labels, 
$\vec{\alpha}$ are hyperparameters and
$B(\vec{\alpha})$ is the multivariate beta function. For binary segmentation ($K=2$), the Dirichlet distribution reduces to the beta distribution 
\be
\label{Beta}
D \left( p_1, p_2 | (\alpha, \beta) \right) = \frac{1}{B(\alpha, \beta)} p_1^{\alpha - 1} p_2^{\beta -1}, \,\,\, p_2 = 1- p_1, \,\,\,
B(\alpha, \beta ) = \frac{ \Gamma(\alpha) \Gamma(\beta)}{\Gamma (\alpha + \beta )},
\ee
where we have defined $\vec{\alpha} = ( \alpha, \beta )$. Since $p_j$ is a random variable of the Dirichlet distribution, we can compute the expectation values of their moments. 
We define the aleatoric and epistemic uncertainties associated with each class label $i$ in this evidential distribution as follows. 
\bea
\label{Ualea}
u_a &=& \mathbb{E}_{Diri} [ \text{Var}_{bernoulli} ]  = \mathbb{E}_{Diri} \left( p_i (1-p_i) \right), \\
\label{Uepis}
u_e &=& 
\text{Var}_{Diri} [ \text{E}_{bernoulli} ] = \text{Var}_{Diri} \left( p_i \right).
\eea
Using the integral identity 
$$
\int^1_0 dx\,\, x^a\, (1-x)^b = 
\frac{\Gamma (1+a)\Gamma (1+b)}{\Gamma (a+b+2)},
$$
we can simplify \eqref{Ualea} and \eqref{Uepis} to read 
\be
\label{U_sim}
u_a =
\frac{\alpha_i (S - \alpha_i)}{S(S+1)},
\,\,\,\,\,
u_e =
\frac{\alpha_i (S - \alpha_i)}{S^2(S+1)},
\,\,\,\,\,\,
S = \sum^K_{i=1} \alpha_i.
\ee
For binary segmentation ($K=2, \vec{\alpha} \equiv (\alpha, \beta)$), these expressions
reduce to 
\be
\label{U_binary}
u_a =
\frac{\alpha \beta}{S(S+1)},
\,\,\,\,\,
u_e =
\frac{\alpha \beta}{S^2(S+1)},
\,\,\,\,\,
S = \alpha + \beta,
\ee
where the foreground and background labels share identical uncertainties with $u_a, u_e$ symmetric under the exchange $\alpha \leftrightarrow \beta$. 
Now in the original formulation of EDL by Sensoy et al.~\cite{sensoy}, the authors alluded to 
a different notion of uncertainty called `Dempster
uncertainty' that is defined within the framework 
of Dempster-Shafer-Theory of Evidence~\cite{Dempster} 
and the formalism of Subjective Logic~\cite{SL}. 
This notion of uncertainty is commonly invoked in works related to EDL. 

The Dempster–Shafer Theory of Evidence (DST) is a generalization of the Bayesian theory to subjective probabilities~\cite{Dempster}. It assigns `belief functions' and an overall uncertainty $u_d$ to the set of exclusive possible states, e.g., in our context, the set of possible class labels for pixels of a medical image. These quantities which measure the reliability of the model 
satisfy the following sum rule
\be
\label{DSTrule}
u_d + \sum_{j=1}^K b_j = 1, \,\,\,\,\, b_j = \frac{e_j}{S},\,\,\,\,\, S = \sum_{j=1}^K (e_j + 1 ),
\ee
where $e_j$ is the `evidence function' derived for class $j$, $u_d$ is the overall Dempster uncertainty and
$b_j$ are the `belief functions' that correspond to a subjective opinion measuring the amount of 
data support in favor of an input to be classified into a definite class. 
The uncertainty $u_d$ and evidence functions are connected
to the parameters of the Dirichlet distribution $\vec{\alpha}$ as follows.
\be
\label{dempster_overall}
e_j = \alpha_j - 1, \,\,\,\,\, 
u_d = \frac{K}{\sum_{i=1}^K \alpha_i } =
\frac{K}{S}.
\ee
Further, eqn. \eqref{dempster_overall} implies that we can relate the Dempster uncertainty $u_d$ to the aleatoric and epistemic uncertainties $u_a, u_e$ as follows.
\be
\label{simple_demp}
u_d = \frac{K u_e}{u_a}.
\ee
Thus, up to an overall scaling factor $K$ which is the number of class labels, the Dempster uncertainty $u_d$ is the epistemic uncertainty $u_e$ in units of the aleatoric uncertainty $u_a$. In Subjective Logic \cite{SL}, 
the Dempster uncertainty $u_d$ represents the overall degree of vacuity of evidence with respect to the Dirichlet distribution. 
As reviewed in e.g. \cite{kendall}, the aleatoric uncertainty $u_a$ is representative of the stochasticity attributable to the 
image data uncertainty/noise. On the other hand, epistemic uncertainty is more of an uncertainty related to the limitation of the underlying neural network. This limitation can be due to an underlying inadequacy of the model's structure with respect to the data distribution and/or some fundamental insufficiency of the training data. There are thus different notions of uncertainties $\{ u_d,u_a,u_e \}$ at a more granular level as explained above.

When implemented on a neural network, as originally proposed in \cite{sensoy}, the model's final output variables are the
evidence variables 
\be
\label{evid_var}
e_i = \alpha_i - 1 = f_{model}(\vec{x}_i | \vec{w} ),
\ee
where $f_{model}$ denotes the output function of the neural network, $\vec{w}$ denotes the weight vectors of the neural network, and  $\vec{x}_i$ denotes the model's inputs,  
Once $\vec{e}$
and thus $\vec{\alpha}$
are obtained, we can make an inference of the pixel's class through the model's prediction of class probabilities.
Taking expectation values with respect to the Dirichlet distribution (whose parameters $\vec{\alpha}$ have been learned), the mean class probabilities are 
\be
\label{meanP}
\langle p_k \rangle = \int^1_0 dp_k\, p_k  \, \left(
\frac{1}{B(\vec{\alpha})} \prod_{i=1}^K p_i^{(\alpha_i - 1)}
\right)= \frac{\alpha_k}{\sum^K_{i=1} \alpha_i} \equiv \frac{\alpha_k}{S}, \,\,\,
k = 1,2,\ldots, K.
\ee

%\subsection{Related Work}

\section{Methodology}

\subsection{On datasets, model structure and loss function}

Open-source datasets were employed for our model validations.  In this work, we chose to use two MRI image datasets drawn from the Medical Segmentation Decathlon \cite{Decathlon}: cardiac (left atrium) mono-modal MRI and prostate (combining central gland and peripheral zone) MRI (T2 and apparent diffusion coefficient (ADC) modes), both of which were 
described in \cite{Decathlon} to display large inter-subject variability. The ground truth images were annotated by medical experts in 
Radboud University, Nijmegen Medical Centre for the prostate dataset and those in King's College London for cardiac dataset \cite{Decathlon}.

For our neural network architecture, we adopted a 2D U-Net backbone with a 2-channel convolution layer as the final layer, the output variables being the evidence variables of eqn. \eqref{evid_var} related to the 
Dirichlet distribution parameters $\alpha, \beta$ 
as follows.
\be
e_1 = \alpha - 1, \,\,\,\,\,
e_2 = \beta - 1, \,\,\,\,\, e_i \geq 0.
\ee
Inference on the binary class of each pixel is determined by the relative magnitude of the mean class probabilities 
\be
\label{classP}
\hat{p}_1 = \frac{\alpha}{S}, \,\,\,\,\,
\hat{p}_2 = \frac{\beta}{S}, \,\,\,\,\, S = \alpha + \beta,
\ee
where $\hat{p}_{1,2}$ refer to the probabilities
of the pixel belonging to background and foreground respectively, these quantities being averaged over the Dirichlet distribution parametrized by $\alpha, \beta$. 
Our U-Net has the following structure:

\begin{itemize}

\item a $320 \times 320$ voxel image with
N = 1 (cardiac) or 2 (prostate) channels;

\item each downsampling level consists of
two convolutional layers each with a $(3 \times 3 )$ kernel, ReLU activation and 
equipped with a dropout unit with rate 0.1, followed by maximum pooling with kernel size $(2\times 2)$; 

\item the downsampling operation is performed 4 times, with
the number of convolutional filters for each layer being $\{32, 64, 128, 256\}$ respectively; 

\item at each of the four upsampling levels, features from the contraction path are concatenated with the corresponding upsampled (bilinear) features;

\item the final 2-channel pointwise convolution layer yields the non-negative evidence variables $e_1, e_2$ which are used to infer pixel class via eqn. \eqref{classP};

\end{itemize}
This model contained about $2\times 10^6$ free weight parameters. We also took it to be the `baseline model' after replacing the final layer's activation function with softmax function and using a Dice loss function for training. 

For EDL model's loss function, among the few types proposed in \cite{sensoy}, we found the Bayes risk function obtained by averaging mean-squared error over the Dirichlet distribution to be most effective. It reads
\be
\label{bayes}
\mathcal{L}_{Bayes} = \int || \vec{p} - \vec{y} ||^2   \frac{1}{B(\vec{\alpha} )} \prod_{l=1}^K  p^{\alpha_l - 1}_l dp
= \sum_{j=1}^K (y_j - \hat{p}_j )^2 + \frac{\hat{p}_j (1 - \hat{p}_j )}{S+1}, 
\ee
where $\hat{p}_j$ are the model's probabilistic prediction for the pixel's label defined in eqn. \eqref{classP} earlier. This also needs to be 
augmented by a regularizer-type loss term that 
shifts the Dirichlet distributions for incorrectly-labeled pixels towards uniform distributions so that the errors are characterized by higher uncertainties. As formulated in \cite{sensoy}, it reads
\be
\label{KL_term}
\mathcal{L}_{KL} = \log \left( \frac{\Gamma (\sum_j \tilde{\alpha}_j )}{\Gamma(K) 
\prod_j \Gamma (\tilde{\alpha}_j)}  \right) + \sum_{j=1}^K ( \tilde{\alpha}_j -1 ) 
\left[ 
\psi( \tilde{\alpha}_j ) - \psi (\sum^K_j \tilde{\alpha}_j )
\right],
\ee
where $\tilde{\alpha}$ are the $\alpha$'s of the incorrectly labeled pixels. This loss term
is proportional to the Kullback-Leibler divergence (see e.g. \cite{cover}) between the Dirichlet and uniform distribution. In many or our ablation experiments, the sum of the loss terms \eqref{bayes} and \eqref{KL_term} led to an effective segmentation model with good uncertainty-error correlations, but we found that 
for many choices of hyperparameters, it attained Dice coefficients which were markedly inferior to the baseline U-Net. This deficiency was resolved 
after adding another focal Dice loss term defined as follows. 
\be
\label{focal}
\mathcal{L}_{Dice} =  1 - \left( 
\frac{2  \frac{1}{N} \sum^N_{i=1}  y^i_t y^i_p  }{ \frac{1}{N} \sum^N_{i=1}  (y^i_t + y^i_p)}
\right)^3,
\ee
where $y^{i}_t, y^i_p$ 
denote the ground truth label and model's prediction for pixel $i$ and $N$ is the total number of pixels.
The final form of our loss function is the sum
\be
\label{Loss_sum}
\mathcal{L} =  \mathcal{L}_{Bayes}
 + \lambda_{KL} \mathcal{L}_{KL} + \lambda_{Dice}
 \mathcal{L}_{Dice},
\ee
where $\lambda$'s are hyperparameters of which optimal values we found to be
$
\lambda_{KL} = 0.20, \lambda_{Dice} = 0.15
$
for cardiac dataset and 
$
\lambda_{KL} = 0.20, \lambda_{Dice} = 0.01,
$
for prostate dataset.\footnote{The optimal coefficients were found with validation data arising from a 4:1 split of the training dataset, excluding a hold-out testing set that comprises of $20\%$ of the entire dataset. The higher $\lambda_{KL}/\lambda_{Dice}$ ratio required for the prostate (vs cardiac) dataset may be related to it presenting a higher level of difficulty for segmentation as reflected in Table \ref{tab_results_prostate}.}
Towards comparing the EDL model against more common approaches of uncertainty quantification in literature, we implemented a Deep Ensemble model and MC Dropout model for the same datasets. 
Each of the 5 neural networks used in the Deep Ensemble model share the same U-Net architecture with random weight initialization performed via the \emph{He}-uniform initializer \cite{He} to induce different initial conditions.
For the
MC Dropout model, the same base model
was used for passing 30 forward passes with the dropout layers activated for model prediction. 
All models were trained on a NVIDIA A100 GPU 
for 800 epochs with a learning rate of $10^{-4}$ using Adam optimizer (implemented in Tensorflow).

\subsection{On active learning}

Towards the goal of incorporating EDL in real-time clinical workflow which often involves evolving large-scale datasets, we also performed a preliminary study of whether/how the predicted uncertainties can be useful for active learning. 
  
Let the training dataset be split into a labeled set $\mathcal{D}_L$ and an unlabeled set $\mathcal{D}_U$. At any iteration, only the labeled set is used to train the model.
Each iteration of active learning involves picking a number ($N_u$)
of images $\tilde{X}$ from the unlabeled set $\mathcal{D}_U$ using some `acquisition/query' function
$\mathcal{Q}$, 
\be
\label{query}
\{ \tilde{X}_1, \tilde{X}_2, \ldots, \tilde{X}_{N_u} \} = \mathcal{Q} \left(  \{ X: X \in \mathcal{D}_U \} \right),
\ee
which represents a selection criterion such that 
a small subset of the unlabeled pool
can be added to expand the labeled pool for model training, formally, 
$
\mathcal{D}_U  \rightarrow \mathcal{D}_U \setminus   \{ \tilde{X}_1, \ldots, \tilde{X}_{N_u} \}, \,\,\,\,\,\,
\mathcal{D}_L  \rightarrow \mathcal{D}_L  \cup   \{ \tilde{X}_1, \ldots, \tilde{X}_{N_u} \}
$.
The acquisition function $\mathcal{Q}$ ranges from simply a random choice to more complicated selection strategies. As mentioned in \cite{Less,Tan,Wolfram}, uncertainty sampling has proven to be one of the best active learning technique, with $\mathcal{Q}$ defined as the subset with highest uncertainty. A common measure is the Shannon entropy. In our work here, we explore the effectiveness of using the different uncertainties in EDL : Dempster, epistemic and aleatoric for defining the query function $\mathcal{Q}$.  
\be
\label{Qentropy}
\mathcal{Q} =    \underset{\{ \tilde{X}_1, \ldots \tilde{X}_{N_U}  \} }{ \text{arg max}}\, \sum_{i \in \tilde{X}}
U_i, \,\,\,\,\,\,\,\,\,
U_i \in \{ U_{shannon}, u_d, u_e, u_a \},
\ee
where $i$ is the pixel index, 
$U_{shannon} \equiv - \sum_{j=1}^K p_j \log p_j $, and $u_d, u_e, u_a$ are defined in 
eqns. \eqref{simple_demp} and \eqref{U_sim}. 
The active learning algorithm follows mostly that implemented in \cite{Less,Tan}. For the 
prostate dataset (481 training samples), each active learning round involves 10 training epochs with $N_u = 6$
samples with the highest uncertainties acquired. For the larger cardiac dataset (1817 training samples), we took each iteration to involve 20 training epochs with $N_u = 20$. With these hyperparameters, 
the validation Dice coefficient converged at the $50^{\text{th}}$ iteration. Compatible with results obtained in \cite{Tan}, $0.68$ and $0.75$ of the entire training data were required to attain similar Dice coefficients achieved by using the full training data for the cardiac and prostate datasets respectively.  

\subsection{On measures of uncertainty-error correlations}

A major goal of this work is to scrutinize
the degree of correlation between uncertainty measures and model errors. Among the spectrum of indices we used, two of them are more discriminative and sensitive in detecting such a correlation. The first is the point-biserial coefficient between the dichotomous
variable (true/false label prediction)
and the continuous uncertainty measure. 
The other is the Kolmogorov-Smirnov statistic (see e.g. \cite{KS})
measuring the discrepancy between the empirical cumulative distribution functions of the two separate sets of true/false labels. For both these coefficients, we computed them for each 2D slice\footnote{The 2D slices are defined along the axial plane for the prostate and the sagittal plane for the cardiac dataset.} of the MRI image since many biomedical applications involving segmentation operate at this scale, and then averaged over the testing dataset to obtain a representative score for each deep learning model. 

For any uncertainty-aware model, 
the Brier score \cite{brier} and negative log likelihoods (NLL) \cite{James} are two commonly used proper scoring rules in Decision Theory to measure how well-calibrated the probabilistic model predictions are. For each 2D slice, we computed the average Brier score, NLL, Dice and accuracy values as well as the average uncertainty measure. The Spearman's rank coefficient between the averaged uncertainty estimates and these various scores were then collected to probe whether 2D-image-averaged uncertainties correlated with the degree of calibration of each model.

\section{Results}
\label{sec:results}

\subsection{EDL showed the strongest uncertainty-error correlation}

Overall, our results demonstrated that
EDL models exhibited the most significant correlation between uncertainties and errors, as measured by the point-biserial coefficient and the Kolmogorov-Smirnov statistic differentiating between the eCDFs of correctly and falsely labeled voxels. 
Detailed numerical results are collected 
in Table \ref{tab_results_prostate} for the both datasets, with the highlighted cells showing the most relevant ones for the strongest uncertainty-error correlations.

In terms of the Dice coefficient, EDL achieved the same performance relative to all the other models for both datasets, indicating that segmentation accuracy was not compromised while having a more robust uncertainty quantification. Our experimental results also suggested that it is generally useful to consider the three subtypes of evidential uncertainties as we found 
aleatoric uncertainty to have the highest point-biserial coefficient (0.54) for cardiac dataset, while Dempster and epistemic uncertainties attained the highest value (0.51) for prostate dataset. 

\begin{table}[h!]
\small
\caption{\small Collection of correlation indices and Dice coefficients.  \textbf{Abbrev}: Demp, Epis, Alea: Dempster, epistemic and aleatoric uncertainties, NLL: negative log-likelihood, 
K.S.:Kolmogorov-Smirnov statistic, 
Pt-Biserial: point-biserial coefficient between error and uncertainty, Spearman(...): Spearman's coefficient between uncertainty (U) and (i)Brier Score (B),(ii)negative log-likelihood (NLL), (iii)Dice coefficient (D),(iv)Accuracy (A). }
\label{tab_results_prostate}
\begin{tabular}{|c|c|c|c|c|c|c|}
\hline
\hline
\textcolor{blue}{Prostate Dataset} & \multicolumn{3}{c|}{ \textbf{EDL}} & 
{\bfseries Baseline} & 
{\bfseries MC Dropout} & 
{\bfseries Deep Ensemble} \\
\hline
$\,$ & \cellcolor{gray!30} Demp & \cellcolor{gray!30} Epis & Alea & Entropy & [Std.Dev.] & [Std.Dev.] \\
\hline
{\bfseries Dice Coef.} & \multicolumn{3}{c|}{$0.81$} & $0.80$ &
$0.80$ & $0.80$ \\
\hline
{\bfseries Brier Score} & \multicolumn{3}{c|}{$3.8 \times 10^{-3}$} & $4.0 \times 10^{-3}$ &
$3.4 \times 10^{-3}$ & $2.9 \times 10^{-3}$ \\
\hline
{\bfseries NLL} & \multicolumn{3}{c|}{$0.031$} & $0.055$ &
$0.035$ & $0.023$ \\
\hline
{\bfseries K.S. statistic} & \multicolumn{3}{c|}{\cellcolor{gray!30} $0.94$} & $0.62$ &
$0.56$ & $0.68$ \\
\hline
{\bfseries Pt-Biserial} & \cellcolor{gray!30} 0.51 &\cellcolor{gray!30}  0.51 & 0.48 & 0.16 & 0.38 & 0.45 \\
\hline
{\bfseries Spearman(B,U)} & 0.89 &  0.90 & 0.87 & 0.73 & 0.77 & 0.77 \\
\hline
{\bfseries Spearman(NLL,U)} & 0.90 &  0.90 & 0.89 & 0.68 & 0.67 & 0.59 \\
\hline
{\bfseries Spearman(D,U)} & -0.42 &  -0.45 & -0.38 & -0.44 & -0.53 & -0.62 \\
\hline
{\bfseries Spearman(A,U)} & -0.77 &  -0.79 & -0.75 & -0.73 & -0.80 & -0.77 \\
\hline
\hline
\textcolor{blue}{Cardiac Dataset} & \multicolumn{3}{c|}{$\,$} & 
$\,$ & 
$\,$ & 
$\,$ \\
\hline
$\,$ & Demp& Epis &  \cellcolor{gray!30} Alea & Entropy & [Std.Dev.] & [Std.Dev.] \\
\hline
{\bfseries Dice Coef.} & \multicolumn{3}{c|}{$0.95$} & $0.94$ &
$0.94$ & $0.94$ \\
\hline
{\bfseries Brier Score} & \multicolumn{3}{c|}{$1.8 \times 10^{-4}$} & $2.4 \times 10^{-4}$ &
$1.9 \times 10^{-4}$ & $1.6 \times 10^{-4}$ \\
\hline
{\bfseries NLL} & \multicolumn{3}{c|}{$0.0051$} & $0.0034$ &
$0.0018$ & $0.0012$ \\
\hline
{\bfseries K.S. statistic} & \multicolumn{3}{c|}{\cellcolor{gray!30} $0.98$} & $0.59$ &
$0.67$ & $0.76$ \\
\hline
{\bfseries Pt-Biserial} & 0.53 & 0.51 & \cellcolor{gray!30} 0.54 & 0.17 & 0.45 & 0.48 \\
\hline
{\bfseries Spearman(B,U)} & 0.94 & 0.94 & 0.93 & 0.84 & 0.87 & 0.93 \\
\hline
{\bfseries Spearman(NLL,U)} & 0.93 & 0.92 & 0.92 & 0.84 & 0.84 & 0.85 \\
\hline
{\bfseries Spearman(D,U)} & -0.80 & -0.80 & -0.79 & -0.72 & -0.77 & -0.79 \\
\hline
{\bfseries Spearman(A,U)} & -0.90 & -0.91 & -0.90 & -0.82 & -0.90 & -0.89 \\
\hline
\hline
\end{tabular}
\end{table}
In Table \ref{tab_results_prostate}, we have also collected coarse-grained indices which quantify 2D image-averaged correlations. The relevant indices are the Spearman's rank coefficient between mean uncertainty measures and each of Brier score, NLL, Dice and accuracy scores. Results for both datasets indicated that 2D-slice-averaged uncertainty measures of EDL inherited higher correlations with Brier score and NLL relative to the other models, though there was no significant advantage over other methods in terms of their correlations with the Dice coefficient. 

\begin{figure}[h!]
    \centering
    \begin{subfigure}[b]{0.24\textwidth}
        \centering
        \includegraphics[width=\textwidth]{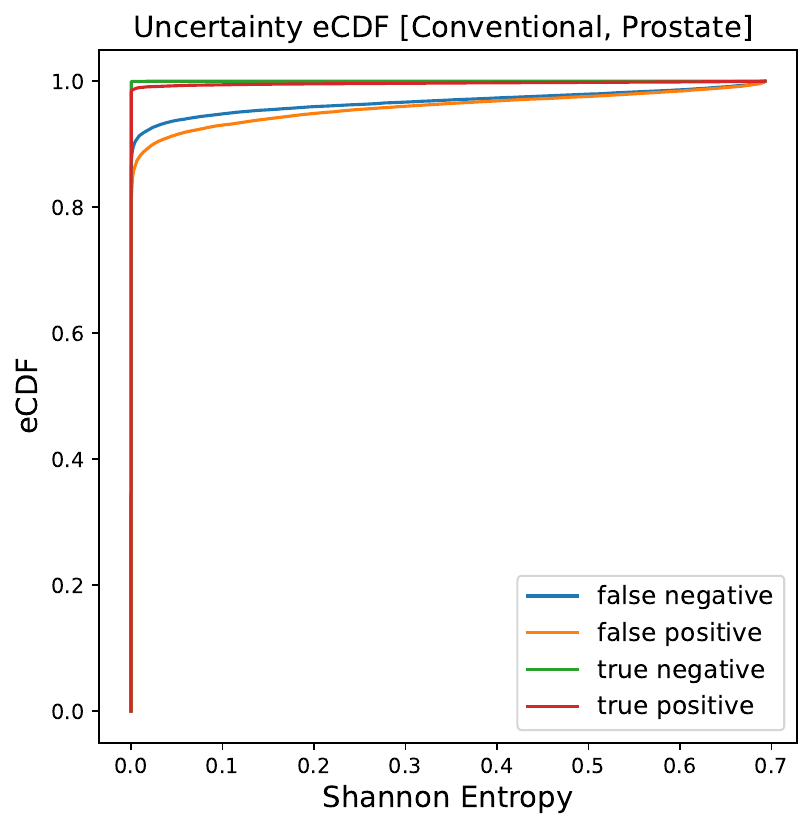}
        \caption{$\,$}
        \label{fig:sub1a}
    \end{subfigure}
    %\hfill
    \begin{subfigure}[b]{0.24\textwidth}
        \centering
        \includegraphics[width=\textwidth]{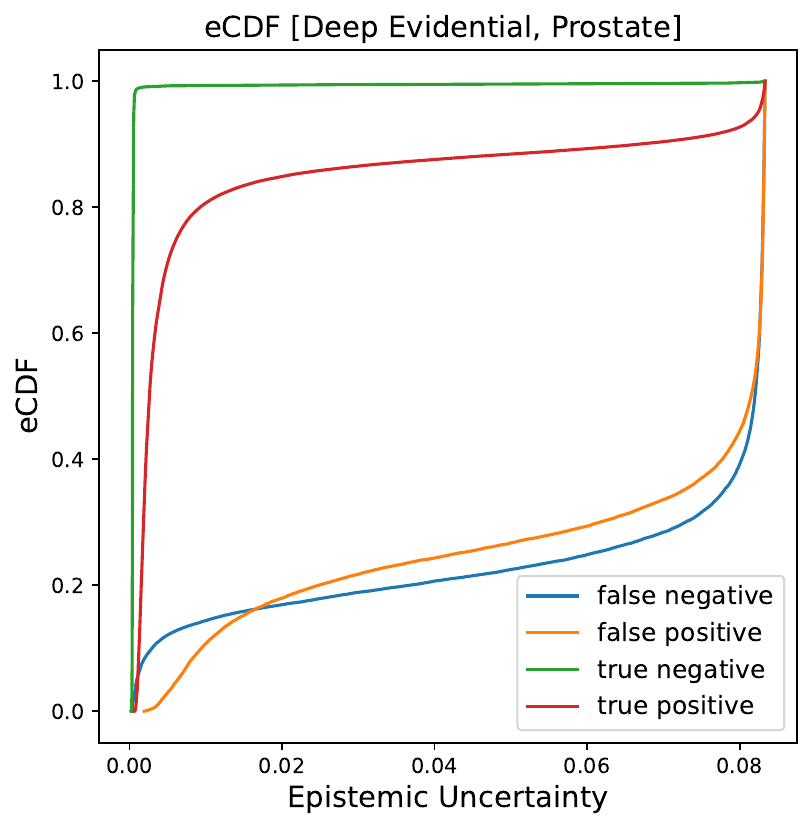}
        \caption{$\,$}
        \label{fig:sub2c}
    \end{subfigure}
    \begin{subfigure}[b]{0.24\textwidth}
        \centering
        \includegraphics[width=\textwidth]{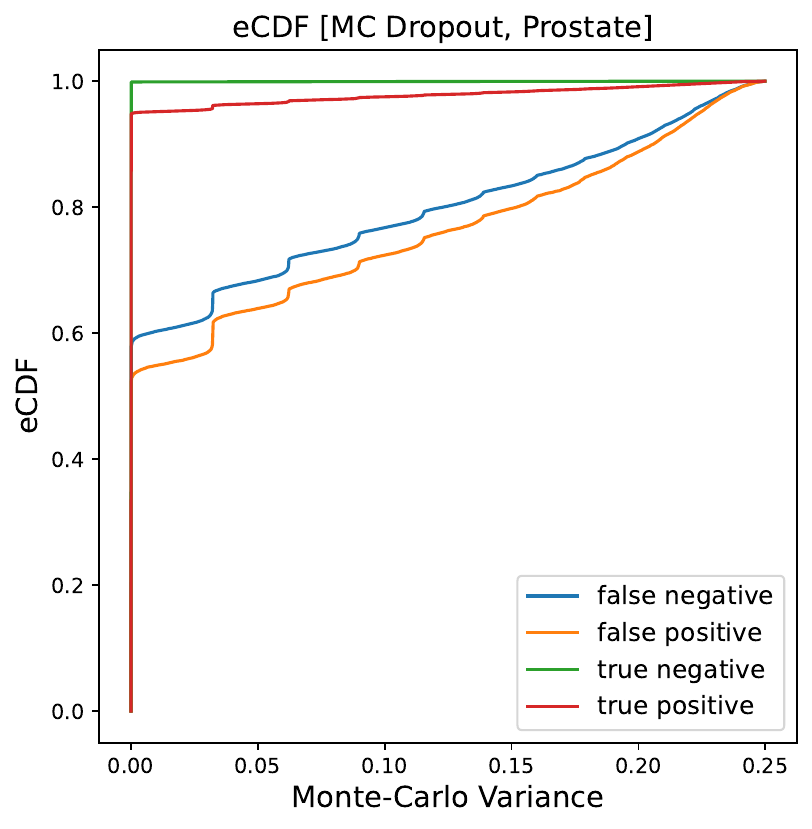}
        \caption{$\,$}
        \label{fig:sub3c}
    \end{subfigure}
    \begin{subfigure}[b]{0.24\textwidth}
        \centering
        \includegraphics[width=\textwidth]{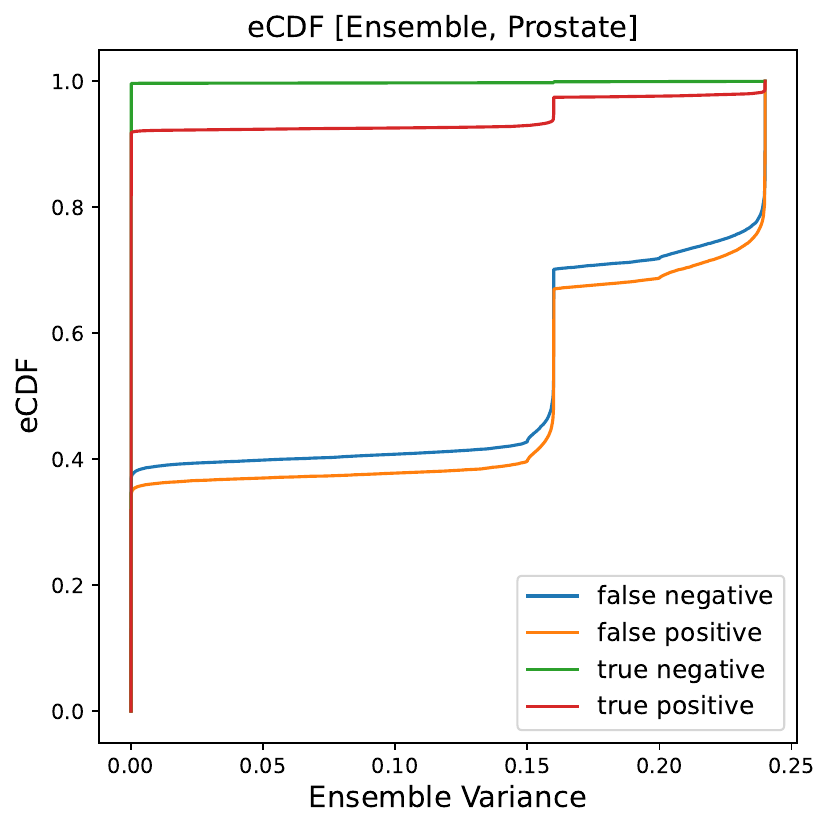}
        \caption{$\,$}
        \label{fig:sub4c}
    \end{subfigure}
    \caption{\small Empirical cumulative distribution 
    functions (eCDF) of uncertainties in various models trained for prostate segmentation. Highly similar trends were also observed for eCDFs pertaining to the cardiac (left atrium) segmentation. }
    \label{fig:ecdf_prostate}
\end{figure}

For a more holistic view on the uncertainty distributions, we examined plots of the empirical cumulative distribution functions (eCDF) of each element of the `confusion matrix': True Positive, True Negative, False Positive, False Negative. In Figure \ref{fig:ecdf_prostate}, one can discern the eCDFs for the correctly labeled voxels (True Positive and Negative) from the falsely labeled ones (False Positive and Negative), with the segregation most apparent for the EDL model. Formally, although simple visual inspection of Fig. \ref{fig:ecdf_prostate} gave us sufficient insight, we can invoke the Kolmogorov-Smirnov statistic 
as a numerical representative index. We found that for both datasets, this statistic was significantly higher for EDL relative to all other methods. Collectively, our various results demonstrated that among all models considered here, EDL showed the strongest uncertainty-error correlation.

%------------uncertainty heatmaps--------------
\subsection{On uncertainty heatmaps}

\begin{figure}[htbp]
    \centering
    \begin{subfigure}[b]{0.45\textwidth}
        \centering
        \includegraphics[width=\textwidth]{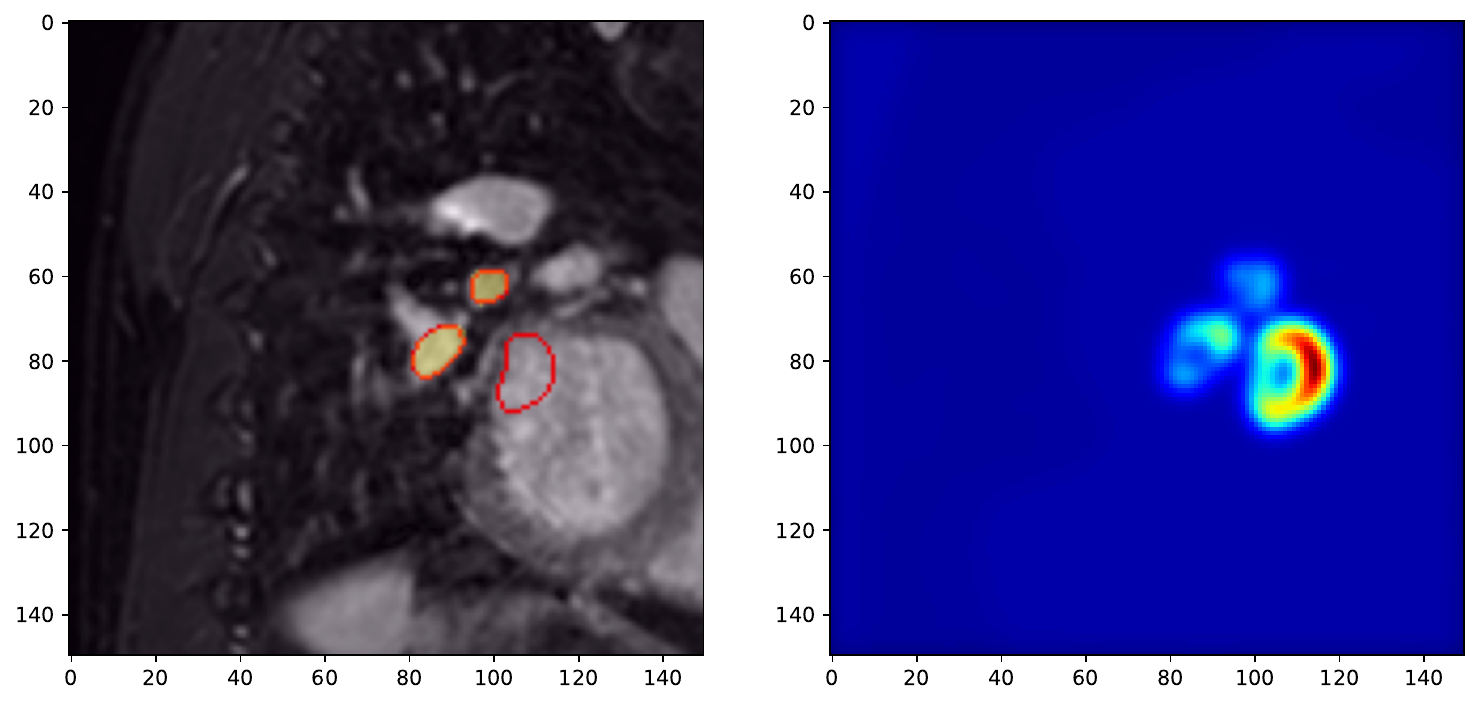}
        \caption{$\,$}
        \label{fig:sub1b}
    \end{subfigure}
    %\hfill
    \begin{subfigure}[b]{0.45\textwidth}
        \centering
        \includegraphics[width=\textwidth]{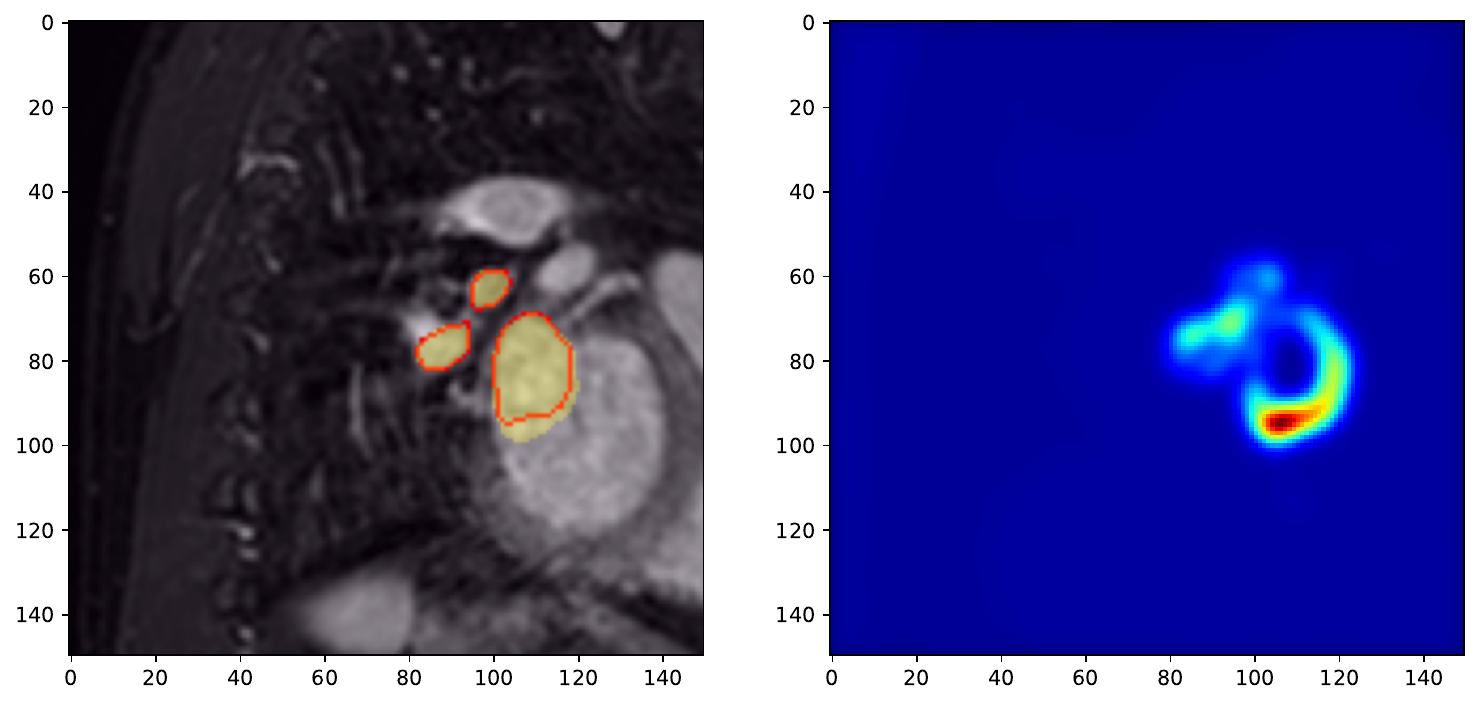}
        \caption{$\,$}
        \label{fig:sub2a}
    \end{subfigure}
    \begin{subfigure}[b]{0.45\textwidth}
        \centering
        \includegraphics[width=\textwidth]{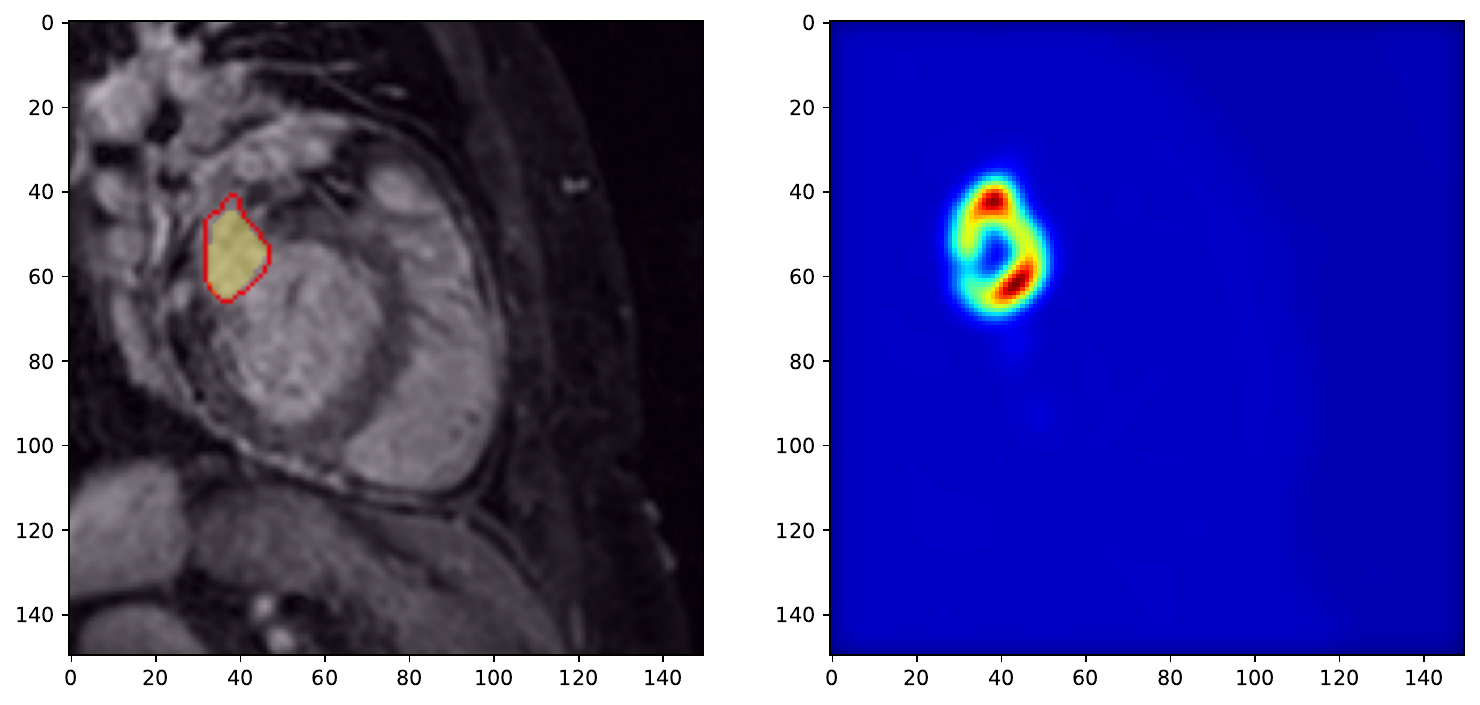}
        \caption{$\,$}
        \label{fig:sub3a}
    \end{subfigure}
    \begin{subfigure}[b]{0.45\textwidth}
        \centering
        \includegraphics[width=\textwidth]{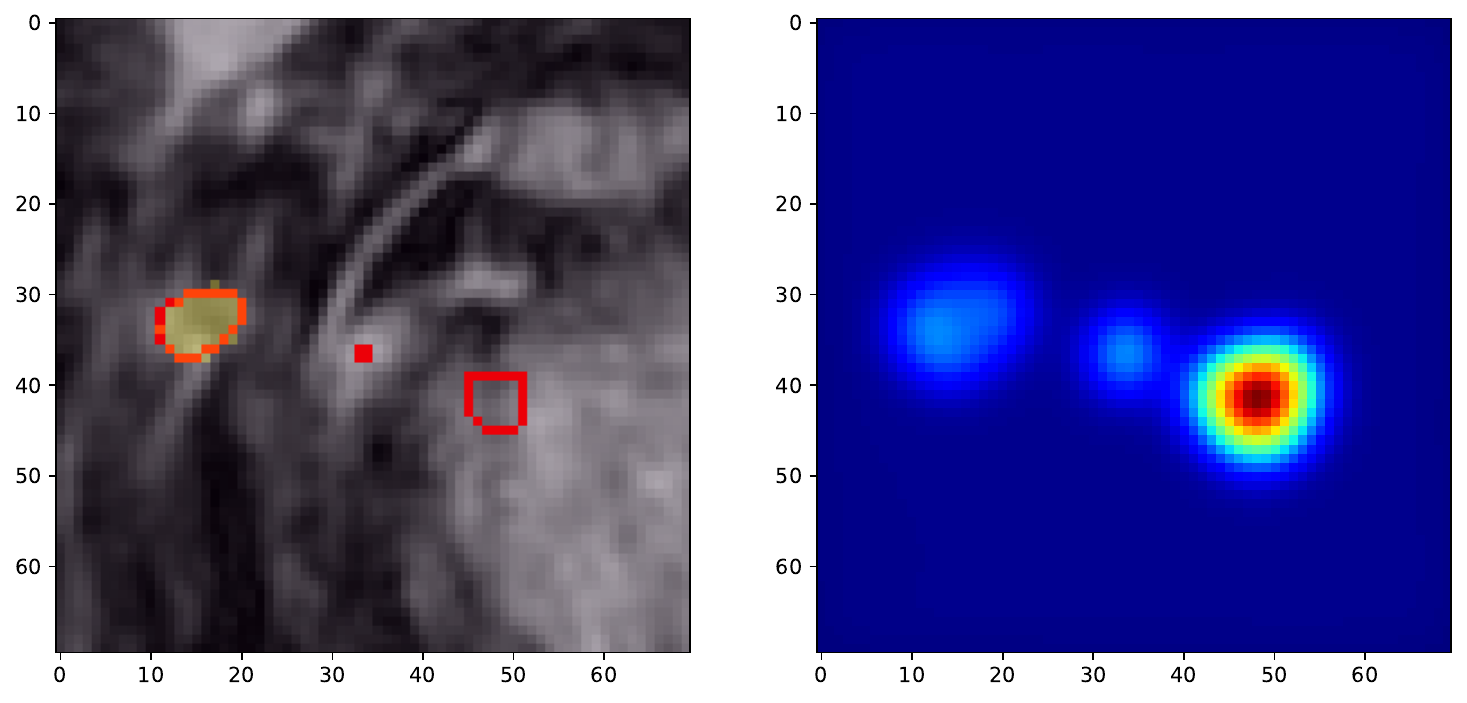}
        \caption{$\,$}
        \label{fig:sub4a}
    \end{subfigure}
    \caption{\small Aleatoric uncertainty heatmaps for segmentation of the left atrium in sagittal T1-weighted cardiac MRI images. Red curves pertain to model-predicted contours overlaid on ground truth masks in yellow. }  \label{fig:uncertainty_heatmaps_cardiac}
\end{figure}

\begin{figure}[htbp]
    \centering
    \begin{subfigure}[b]{0.45\textwidth}
        \centering
     \includegraphics[width=\textwidth]{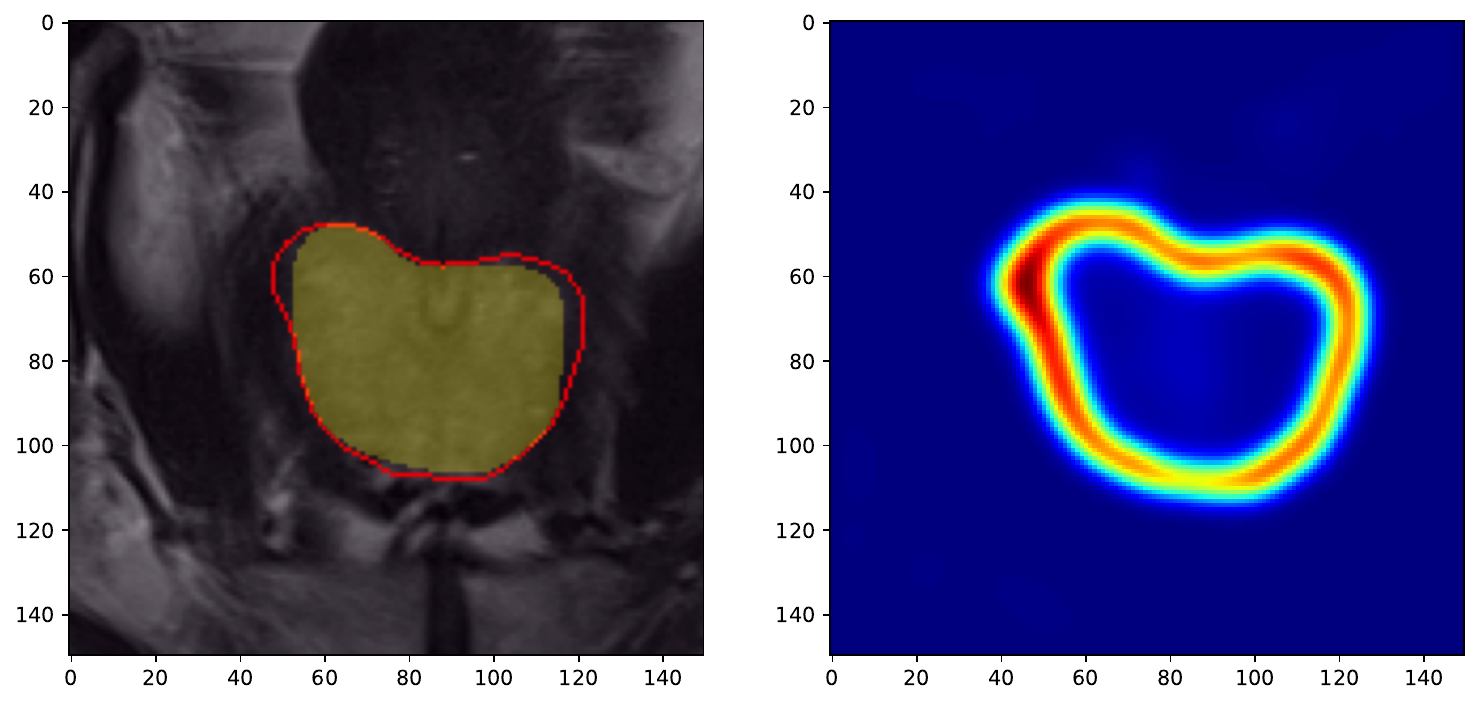}
        \caption{$\,$}
        \label{fig:sub1c}
    \end{subfigure}
    \begin{subfigure}[b]{0.45\textwidth}
        \centering
     \includegraphics[width=\textwidth]{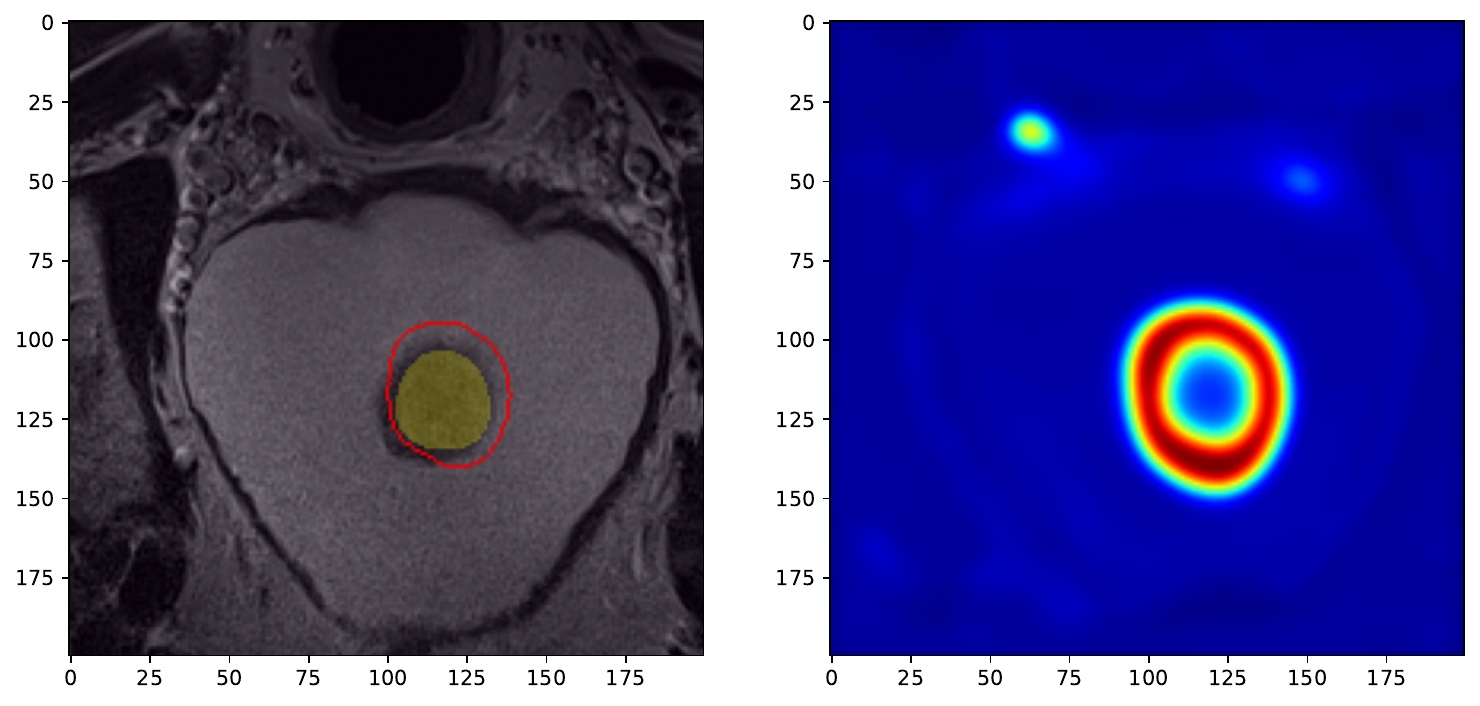}
        \caption{$\,$}
        \label{fig:sub2b}
    \end{subfigure}
    \begin{subfigure}[b]{0.45\textwidth}
        \centering    \includegraphics[width=\textwidth]{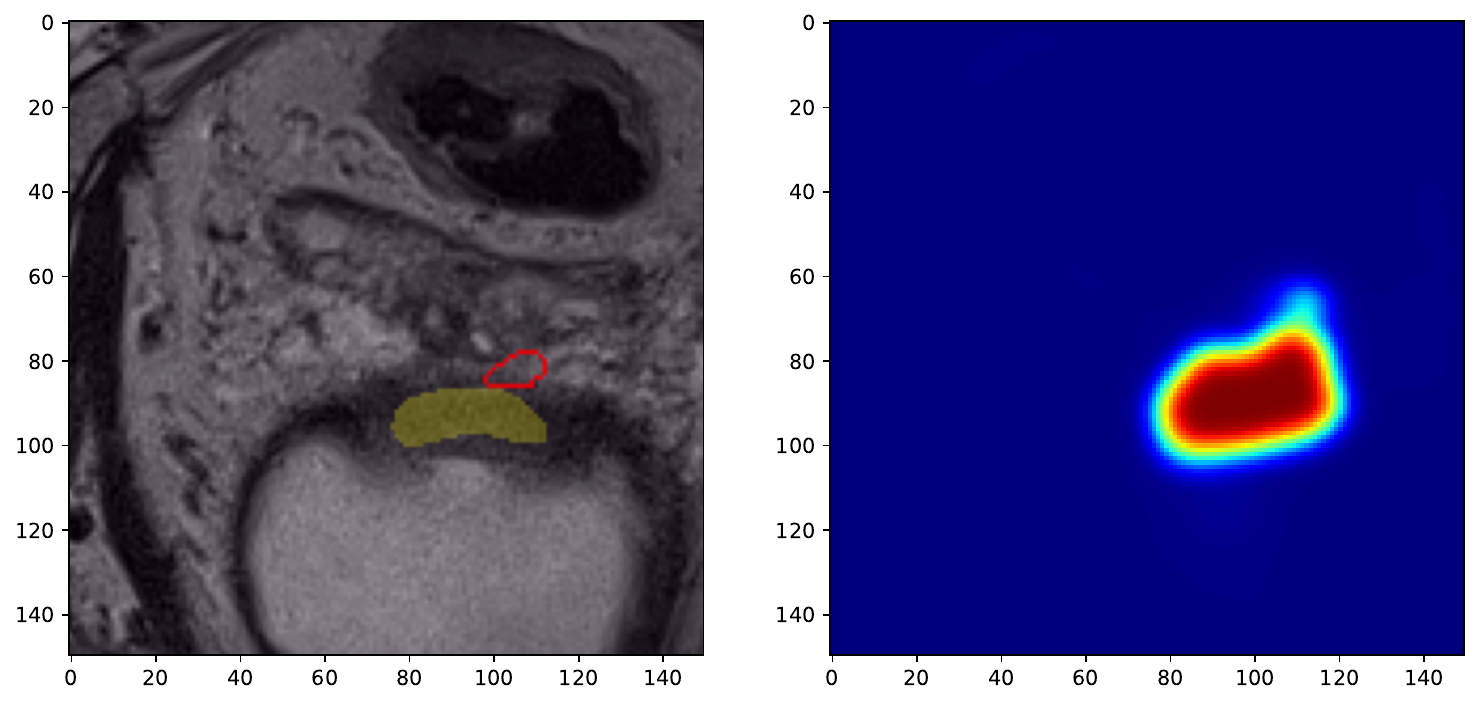}
        \caption{$\,$}
        \label{fig:sub3b}
    \end{subfigure}
    \begin{subfigure}[b]{0.45\textwidth}
        \centering
   \includegraphics[width=\textwidth]{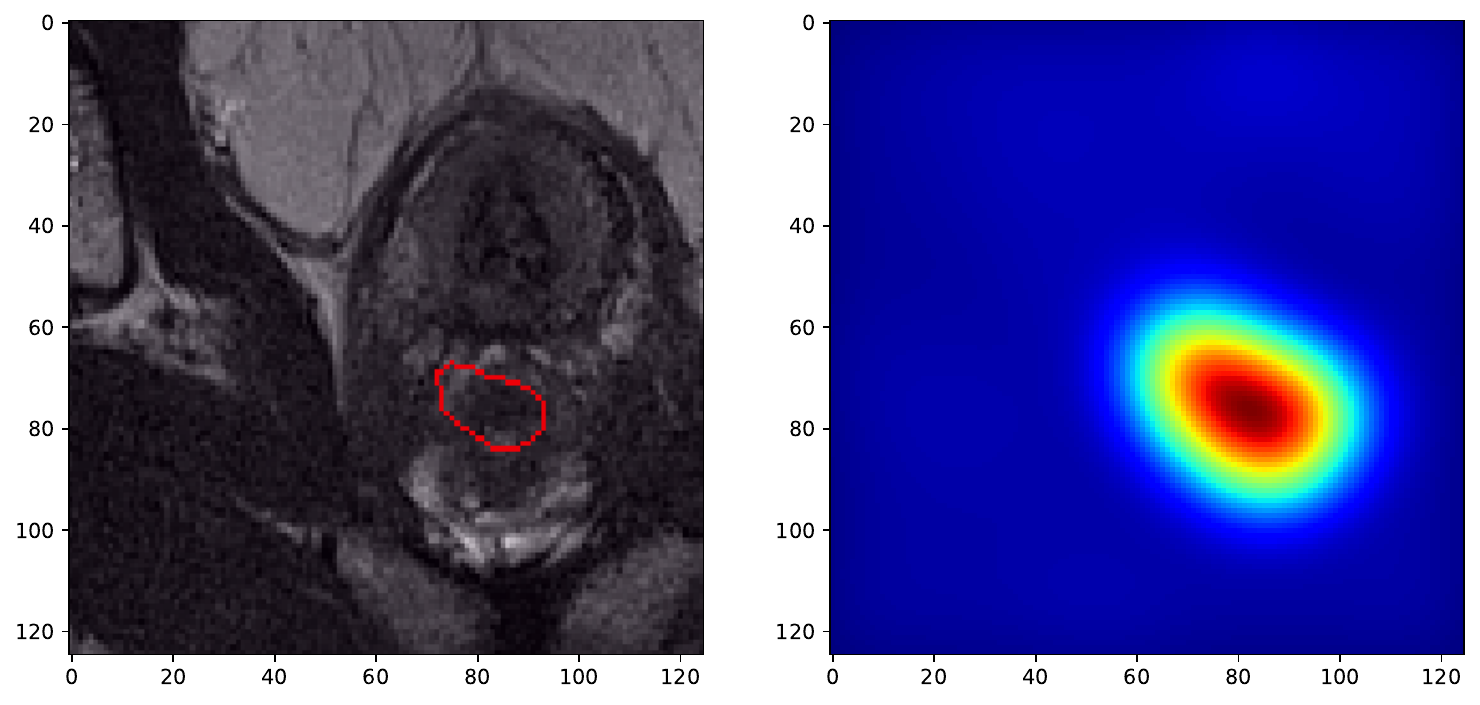}
        \caption{$\,$}
        \label{fig:sub4b}
    \end{subfigure}
    \caption{\small Epistemic uncertainty heatmaps for segmentation of the central and peripheral zones of the prostate in axial T2-weighted MRI images. Note how elevated redder regions of each heatmap appear to mirror deviations between ground truth masks in yellow and their model-predicted contours in red.  }    \label{fig:uncertainty_heatmaps_prostate}
\end{figure}

Visual inspection of the uncertainty heatmaps generated by EDL models suggests that they are useful towards discovering potential regions of model errors. In Figs. 
\ref{fig:uncertainty_heatmaps_cardiac} and
\ref{fig:uncertainty_heatmaps_prostate}, we present a few illustrative images and their uncertainty heatmaps for each dataset as modeled by the EDL framework.  

For the images of the cardiac dataset depicted in Fig.~\ref{fig:uncertainty_heatmaps_cardiac},(b) and (c) are  examples of contours with small model errors in the neighborhood of the boundary curve. We note how discrepancies between the model-predicted (red curve) and ground truth (yellow mask) are often captured by elevated uncertainty regions. (a) and (d) are examples of cases where isolated components of the model-predicted contour contain no foreground pixels; these false-positive areas are marked by elevated uncertainty regions.
For the prostate dataset depicted in Fig.~\ref{fig:uncertainty_heatmaps_prostate},
(a) and (b) are examples of contours with small model errors in the neighborhood of the boundary curve. (c) and (d) are examples of cases where the model-predicted contour misses the ground truth region completely; the elevated uncertainty regions are filled in the interior and extends to actual ground truth regions. 

The basis for an effective uncertainty heatmap is a robust uncertainty-error correlation. Among the different measures, we found that the difference between eCDF plots of true and false labels stood out as the most consistent way of determining the usefulness of uncertainty heatmaps towards discovering model errors. Models which yielded uncertainty heatmaps that mirrored segmentation errors more effectively were always those which showed eCDF trends similar to Fig. \ref{fig:ecdf_prostate}b, with clear geometric deviations between eCDFs of true and false labels. 

%----------------------------------

\subsection{On active learning}

In Table \ref{tab_results_active}, we collect various results for our study of active learning with EDL and a conventional model with uncertainty sampling via Shannon entropy. We found that while attaining similar Dice coefficient as random and uncertainty sampling, EDL-based uncertainty sampling yielded a much higher point-biserial coefficient and Kolomogorov-Smirnov statistic at the end of active learning iterations. Like our previous results using the conventional model training scheme, aleatoric and epistemic uncertainy-based active learning worked best for the cardiac and prostate dataset respectively. We note that all models converged to the optimal Dice coefficient at the end of the 50${}^{\text{th}}$ iteration with 0.68 and 0.75 of the full training dataset size required. Using EDL uncertainty-based sampling did not speed up the convergence (which would have led to greater training dataset reduction). Nonetheless, the much superior uncertainty-error correlation was still inherited by EDL models relative to standard random and uncertainty-based sampling algorithms.  

\begin{table}
\small
\caption{\small Results for active learning algorithms. \textbf{Abbrev}: Demp, Epis, Alea: uncertainty sampling via higher mean Dempster, epistemic and aleatoric uncertainties respectively.
}
\label{tab_results_active}
\begin{tabular}{|c|c|c|c|c|c|c|c|c|c|c|}
\hline
\hline
$\,$ & \multicolumn{5}{c|}{{\bfseries Cardiac dataset}} & \multicolumn{5}{c|}{{\bfseries Prostate dataset}} \\
\hline
 & Demp & Epis& \cellcolor{gray!30} Alea & Random & Entropy &  Demp & \cellcolor{gray!30} Epis  & Alea & Random & Entropy \\
\hline
{\bfseries Dice Coef.} & 
0.94 & 0.94 & 0.94 & 0.93 & 0.94 & 0.81 & 0.81 &
0.80 & 0.80 & 0.80 \\
\hline
{\bfseries Pt-Biserial} & 
0.43 & 0.39 &  \cellcolor{gray!30} 0.48 & 0.23 & 0.22 & 0.48 & \cellcolor{gray!30} 0.52 &
0.46 & 0.27 & 0.26 \\
\hline
{\bfseries K.S. statistic} & 
\multicolumn{3}{c|}{\cellcolor{gray!30} 0.96} & 0.41 & 0.41 &\multicolumn{3}{c|}{\cellcolor{gray!30} 0.95}  & 0.40 & 0.44 \\
\hline
\end{tabular}
\end{table}

\begin{figure}
\centering
\includegraphics[width=0.9\textwidth]{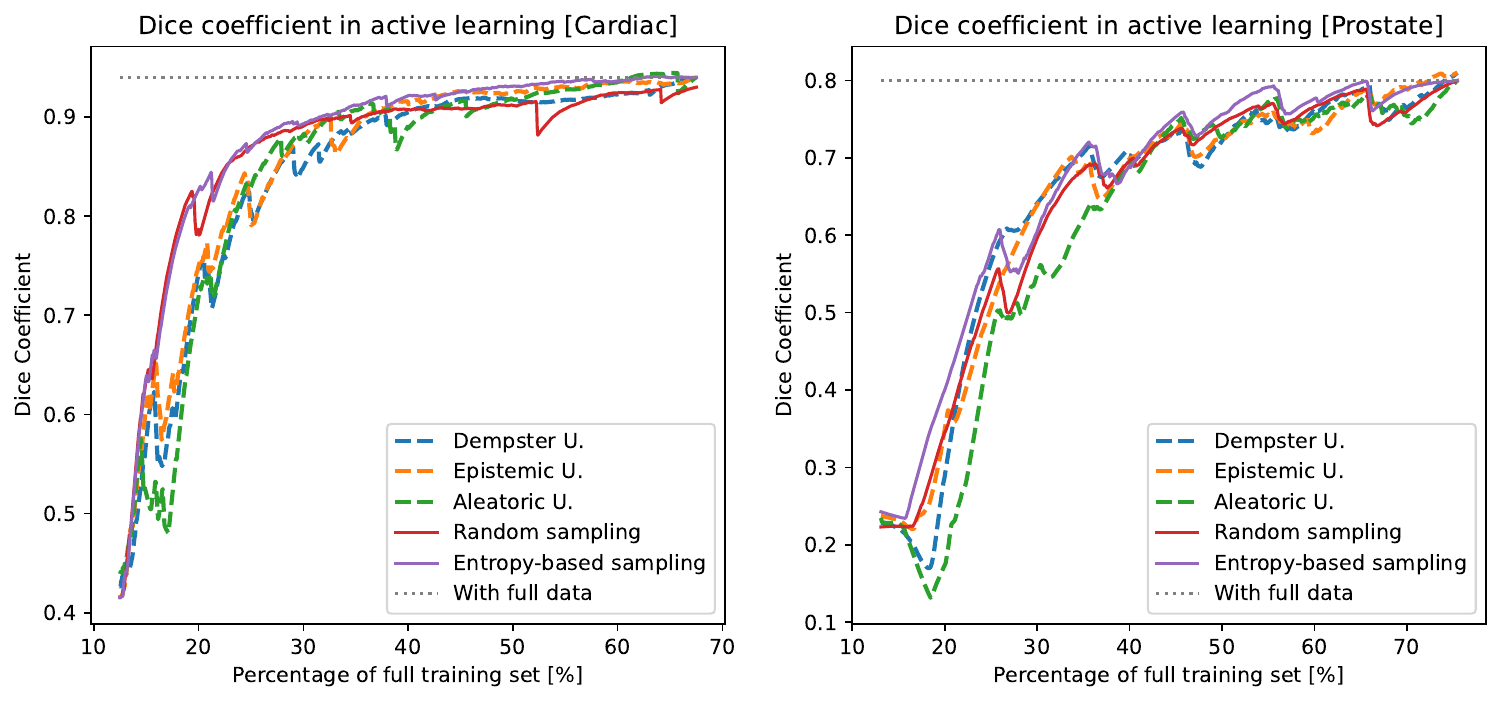}
\caption{\small Evolution of Dice coefficient during active training.  } 
\label{active_dice}
\end{figure}
%\begin{figure}
%\includegraphics[width=\textwidth]{AL_Biserial.pdf}
%\caption{Evolution of Point-Biserial coefficient as a measure of uncertainty-error correlation during active training for both datasets. } 
%\label{active_corr}
%\end{figure}

\section{Discussion}

Collectively, the various results of our work demonstrated that EDL models exhibited the most significant correlation between uncertainties and errors, as measured by the point-biserial coefficient and the Kolmogorov-Smirnov statistic differentiating between the eCDFs of correctly and falsely labeled voxels. 
In terms of the Dice coefficient, EDL attained the same performance relative to all the other models for both datasets, showing that segmentation accuracy was not compromised while having a more robust uncertainty quantification. Our experimental results also suggested that generally, it would be useful to consider the three subtypes of evidential uncertainties for generating uncertainty heatmaps, as we found aleatoric uncertainty to have the highest point-biserial coefficient (0.54) for cardiac dataset, while Dempster and epistemic uncertainties attained the highest value (0.51) for prostate dataset. 
A more global picture of the error-uncertainty relationship can be visualized using the 
eCDFs in each model. As depicted in Fig. \ref{fig:ecdf_prostate}, the EDL models exhibited the most significant deviation between eCDFs of correctly and falsely labeled voxels, and yielded the highest
Kolmogorov-Smirnov statistic parametrizing the differences in eCDFs of uncertainty estimates. 

Ultimately, we would like our model to yield
uncertainty heatmaps that are effective in identifying potential regions of model errors so that the human expert (e.g. dosimetrist) can intervene and rectify the contours whenever necessary. Visual inspection of the uncertainty heatmaps generated by our EDL model 
in Figs. 
\ref{fig:uncertainty_heatmaps_cardiac} and
\ref{fig:uncertainty_heatmaps_prostate}
suggests that they are useful towards realizing this goal.  

Adapting EDL for active learning by using the various notions of uncertainties for defining the query function, we found that while attaining similar Dice coefficient as random and uncertainty sampling, EDL-based uncertainty sampling yielded a much higher point-biserial coefficient and Kolomogorov-Smirnov statistic at the end of active learning iterations. Like our previous results using the conventional model training scheme, aleatoric and epistemic uncertainty-based active learning worked best for the cardiac and prostate dataset respectively.

\section{Conclusion}

Using the cardiac and prostate MRI images of the Medical Segmentation Decathlon \cite{Decathlon} for validation, we found that Evidential Deep Learning (EDL) models with U-Net backbones yielded superior correlations between prediction errors and uncertainties relative to the conventional baseline equipped with Shannon entropy measure, MC Dropout and Deep Ensemble methods. We also examined EDL's effectiveness in active learning, finding that relative to the standard entropy-based uncertainty sampling, they yielded higher point-biserial correlations while attaining similar performances in Dice coefficients. Our spectrum of results furnished novel contextual evidence that EDL models supersede MC Dropout and Deep Ensemble methods in terms of uncertainty-error correlation, and they
yield uncertainty heatmaps that are useful for indicating potential regions of model errors. These superior features of EDL models render them particularly suitable as uncertainty-aware models for segmentation tasks that warrant a higher sensitivity in detecting large model errors.

\bibliographystyle{splncs04}
\bibliography{references}

%Author, F.: Article title. Journal \textbf{2}(5), 99--110 (2016)

%Author, F., Author, S.: Title of a proceedings paper. In: Editor,
%F., Editor, S. (eds.) CONFERENCE 2016, LNCS, vol. 9999, pp. 1--13.
%Springer, Heidelberg (2016). \doi{10.10007/1234567890}

%Author, F., Author, S., Author, T.: Book title. 2nd edn. Publisher,

%Author, A.-B.: Contribution title. In: 9th International Proceedings
%on Proceedings, pp. 1--2. Publisher, Location (2010)

%LNCS Homepage, %\url{http://www.springer.com/lncs}, %last accessed 2023/10/25
%\end{thebibliography}
\end{document}